\documentclass[aip,showpacs,superscriptaddress,onecolumn,preprint]{revtex4}%
\usepackage{amsfonts}
\usepackage{amsmath}
\usepackage{amssymb}
\usepackage{graphicx}%
\begin{document}
\title{First-principles calculations of shocked fluid helium in partially ionized region}
\author{Cong Wang}
\affiliation{LCP, Institute of Applied Physics and Computational Mathematics, P.O. Box
8009, Beijing 100088, People's Republic of China}
\author{Xian-Tu He}
\affiliation{LCP, Institute of Applied Physics and Computational Mathematics, P.O. Box
8009, Beijing 100088, People's Republic of China}
\affiliation{Center for Applied Physics and Technology, Peking University, Beijing 100871,
People's Republic of China}
\author{Ping Zhang}
\thanks{Corresponding author: zhang\underline{ }ping@iapcm.ac.cn}
\affiliation{LCP, Institute of Applied Physics and Computational Mathematics, P.O. Box
8009, Beijing 100088, People's Republic of China}
\affiliation{Center for Applied Physics and Technology, Peking University, Beijing 100871,
People's Republic of China}

\begin{abstract}
Quantum molecular dynamic simulations have been employed to study the equation
of state (EOS) of fluid helium under shock compressions. The principal
Hugoniot is determined from EOS, where corrections from atomic ionization are
added onto the calculated data. Our simulation results indicate that principal
Hugoniot shows good agreement with gas gun and laser driven experiments, and
maximum compression ratio of 5.16 is reached at 106 GPa.

\end{abstract}

\pacs{62.50.-p, 71.30.+h, 31.15.xv}
\maketitle

\section{INTRODUCTION}

\label{sec-introduction}

High-pressure introduced response of materials, which requires accurate
understandings of the thermophysical properties into new and complex region,
has gained much scientific interest recently \cite{PBX:Ernstorfer:2009}. The
relative high temperature and high density are usually referred as the
so-called \textquotedblleft warm dense matter\textquotedblright\ (WDM)- a
strongly correlated state, where simultaneous dissociations, ionizations, and
degenerations make modelling of the dynamical, electrical, and optical
properties of WDM extremely challenging \cite{PBX:Nellis:2006}. WMD, which
provides an active research platform by combining the traditional plasma
physics and condensed matter physics, usually appears in shock or laser heated
targets \cite{PBX:Hicks:2009}, inertial confinement fusion
\cite{PBX:Philippe:2010}, and giant planetary interiors
\cite{PBX:Lorenzen:2009}.

Next to hydrogen, helium is the most abundant element in the universe, and
physical properties of warm dense helium, especially the EOS, are critical for
astrophysics \cite{PBX:Stevenson:1977a,PBX:Stevenson:1977b}. For instance, the
structure and evolution of stars, White Dwarfs, and Giant Planets
\cite{PBX:Saumon:1995,PBX:Ternovoi:2004,PBX:Vorberger:2007,PBX:Perryman:2000},
and therefore the understanding of their formation, depends sensitively on the
EOS of hydrogen and helium at several megabar regime. For all planetary
models, accurate EOS data are essential in solving the hydrostatic equation.
As a consequence, a series of experimental measurements and theoretical
approaches have been applied to investigate the EOS of helium. Liquid helium
was firstly single shocked to 15.6 GPa using two stage light gas gun by Nellis
\emph{et al.}, then double shocked to 56 GPa, and the calculated temperature
are 12000 and 22000 K respectively \cite{PBX:Nellis:1984}. Maximum compression
ratio ($\eta_{max}\mathtt{\approx}6$) was achieved by laser driven shock
experiments with the crossover pressure around 100 GPa \cite{PBX:Eggert:2008},
However, soon after that, Knudson \emph{et al.} have modified $\eta_{max}$ to
be 5.1 \cite{PBX:Knudson:2009}. Theoretically, since the ionization
equilibrium is not interfered with the dissociation equilibrium between
molecules and atoms, helium, which is characterized by monoatomic molecule and
close shell electronic structure, is particularly suitable for the
investigation of the high pressure behavior under extreme conditions. Free
energy based chemical models by Ross \emph{et al.} \cite{PBX:Ross:1986}, Chen
\emph{et al.} \cite{PBX:Chen:2007}, and Kowalski \emph{et al.}
\cite{PBX:Kowalski:2007} have been used to investigate the principal Hugoniot
of liquid helium, and the results are accordant with gas gun experiments.
However, considerable controversies have been raised at megabar pressure
regime, especially since the data was probed by laser shock wave experiments.
Interatomic potential method predict $\eta_{max}$=$4$ for shock compressed
helium \cite{PBX:Ross:1986}, whereas, the EOS used by the astrophysical
community from Saumon \emph{et al.} (SCVH) \cite{PBX:Saumon:1995},
path-integral Monte Carlo (PIMC) \cite{PBX:Militzer:2006}, and activity
expansion (ACTEX) \cite{PBX:Ross:2007} calculations provide an increase in
compressibility at the beginning of ionization. For the initial density of
$\rho_{0}$=$0.1233$ g/cm$^{3}$, SCVH and ACTEX simulation results indicate the
maximum compression ratio lies around 6 at 300 GPa and 100 GPa respectively
\cite{PBX:Saumon:1995,PBX:Ross:2007}, while PIMC calculations suggest
$\eta_{max}$=$5.3$ near 360 GPa \cite{PBX:Militzer:2006}.

On the other hand, quantum molecular dynamic (QMD) simulations, where quantum
effects are considered by the combinations of classical molecular dynamics for
the ions and density functional theory (DFT) for electrons, have already been
proved to be successful in describing thermophysical properties of materials
at complex conditions \cite{PBX:Wang:2010a,PBX:Wang:2010b}. However, the DFT
based molecular dynamic simulations (with or without accounting for excited
electrons) do not provide reasonable results at the ionization region, mainly
because the atomic ionization is not well defined in the framework of DFT.
Considering these facts mentioned above, thus, in the present work, we applied
the corrected QMD simulations to study shock compressed helium, and the EOS,
which is compared with experimental measurements and different theoretical
models, are determined for a wide range of densities and temperatures. The
calculated compression ratio is substantially increased according to the
ionization of atoms in the warm dense fluid.

\section{COMPUTATIONAL METHOD}

\label{sec-method}

The Vienna Ab-initio Simulation Package (VASP)
\cite{PBX:Kresse:1993,PBX:Kresse:1996}, which was developed at the Technical
University of Vienna, has been employed to perform simulations for helium. The
elements of our calculations consist of a series of volume-fixed supercells
including $N$ atoms, which are repeated periodically throughout the space. By
involving Born-Oppenheimer approximation, electrons are quantum mechanically
treated through plane-wave, finite-temperature (FT) DFT
\cite{PBX:Lenosky:2000}, where the electronic states are populated according
to Fermi-Dirac distributions at temperature $T_{e}$. The exchange-correlation
functional is determined by generalized gradient approximation (GGA) with the
parametrization of Perdew-Wang 91 \cite{PBX:Perdew:1991}. The ion-electron
interactions are represented by a projector augmented wave (PAW)
pseudopotential \cite{PBX:Blochl:1994}. Isokinetic ensemble (NVT) is adopted
in present simulations, where the ionic temperature $T_{i}$ is controlled by
No\'{s}e thermostat \cite{PBX:Nose:1984}, and the system is kept in local
equilibrium by setting the electron ($T_{e}$) and ion ($T_{i}$) temperatures
to be equal.

The plane-wave cutoff energy is selected to be 700.0 eV so that the pressure
is converged within 3\% accuracy. $\Gamma$ point is used to sample the
Brillouin zone in molecular dynamics simulations, because EOS can only be
modified within 5\% for the selection of higher number of \textbf{k} points.
64 helium atoms are included in the cubic supercell. The densities selected in
our simulations range from 0.1233 to 0.8 g/cm$^{3}$ and temperatures between 4
and 50000 K, which highlight the regime of the principal Hugoniot. All the
dynamic simulations are lasted for 6000 steps, and the time steps for the
integrations of atomic motion are selected according to different densities
(temperatures) \cite{PBX:timestep}. Then, the subsequent 1000 steps of
smulation are used to calculate EOS as running averages.

\section{RESULTS AND DISCUSSION}

\label{sec-analysis}

\begin{figure}[pt]
\centering
\includegraphics[width=12.0cm]{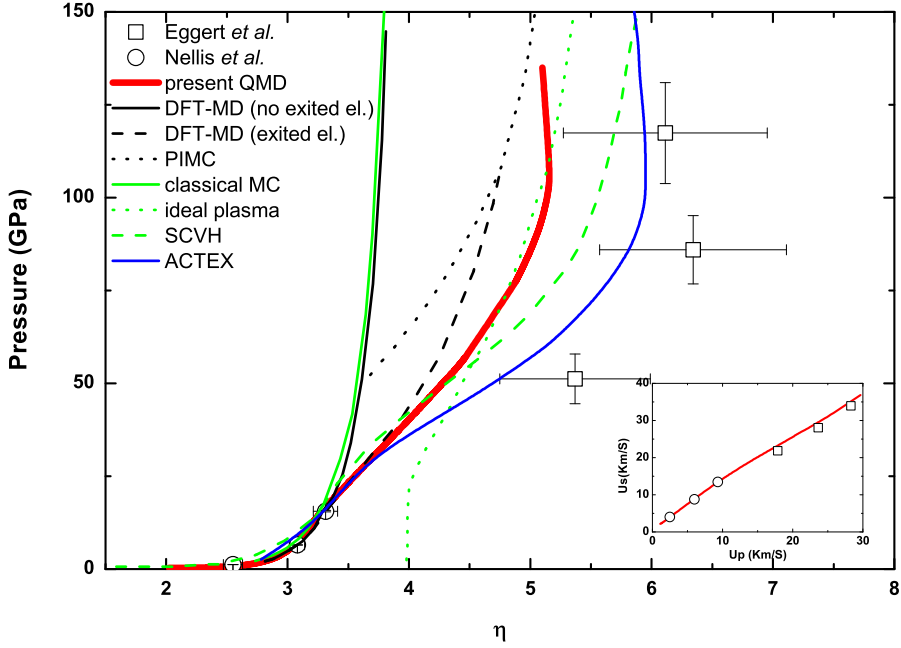}\caption{(Color online) Principal
Hugoniot up to 130 GPa is shown for the present work (red line), where
previous theoretical predictions and experimental measurements are also shown
for comparison. Experiments: Two stage light gas gun data
\cite{PBX:Nellis:1984} is denoted by open circle; Laser driven experimental
results \cite{PBX:Eggert:2008} are labelled by open square. Theories: (i)
Classical MC simulation results \cite{PBX:Aziz:1995} are plotted as green
solid line; (ii) ideal plasma's Hugoniot curve \cite{PBX:Militzer:2006} is
shown as green dotted line; (iii) SCVH method \cite{PBX:Saumon:1995} is
labelled as green dashed line; (iv) ACTEX results \cite{PBX:Ross:2007} are
blue solid line; (v) Previous DFT simulation results \cite{PBX:Militzer:2006}
(with and without accounting for electron excitation) are shown as black
dashed and solid line, respectively; (vi) PIMC results
\cite{PBX:Militzer:2006} are labelled as black dotted line. Inset is the plot
of shock wave velocity verse mass velocity. }%
\label{fig_hugoniot}%
\end{figure}

Different corrections to QMD simulations have already been used to model the
thermophysical properties of WDM, such as the zero point vibrational energy
modification for hydrogen and deuterium \cite{PBX:Holst:2008}. However,
quantitative descriptions of the ionization of atoms in the frame of DFT is
still lacking, except for the Drude model for aluminium introduced by Mazevet
\emph{et al.} \cite{PBX:Mazevet:2005}, but the simple metallic model is not
suitable for studying warm dense helium. Here, we adopt similar approximations
as described in Ref. \cite{PBX:Wang:2010c}, where the effect of atomic
ionization are considered, to study the EOS of fluid helium. In standard
FT-DFT molecular dynamic simulations, the ionization energy is excluded, thus
the EOS should be corrected as \cite{PBX:Wang:2010c}:
\begin{equation}
E=E_{QMD}+N\beta E_{ion},\label{tot_E}%
\end{equation}%
\begin{equation}
P=P_{QMD}+(1+\beta)\frac{\rho k_{B}T}{m_{He}},\label{tot_pressure}%
\end{equation}
where $E_{QMD}$ and $P_{QMD}$ are data obtained from regular QMD calculations,
and $N$ is the total number of atoms considered in supercells. $m_{He}$ and
$k_{B}$ represent the mass of helium atom and Boltzmann constant. The density
and temperature are denoted by $\rho$ and $T$, respectively. $\beta$ stands
for the ionization degree, and $E_{ion}$ is the ionization energy.

\begin{figure}[pt]
\centering
\includegraphics[width=12.0cm]{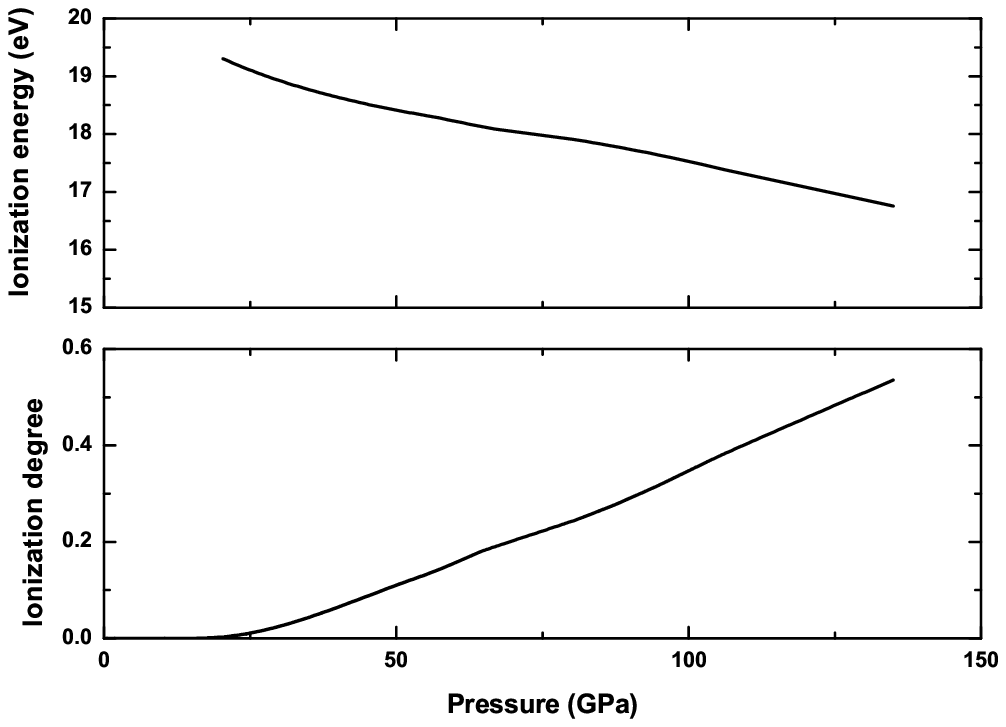}\caption{Ionization energy and
ionization degree are plotted as functions of pressure along the
principal
Hugoniot in the upper and lower panels, respectively.}%
\label{fig_ion}%
\end{figure}

In the present work, only the first ionization level is considered, and the
ionization degree of helium can be evaluated through Saha equation:
\begin{equation}
\label{Saha}\frac{\beta^{2}}{1-\beta}=\frac{2\Omega}{\lambda^{3}}\exp
(-\frac{E_{ion}}{k_{B}T}),
\end{equation}
\begin{equation}
\label{lamda}\lambda=\sqrt{\frac{h^{2}}{2\pi m_{e}k_{B}T}},
\end{equation}
where, $m_{e}$ and $\Omega$ present the mass of electron and volume of the
supercell. The ionization energy $E_{ion}$ is determined by $E_{ion}%
=E^{He+}(\rho,T)-E^{He}(\rho,T)$, where $E^{He+}(\rho,T)$ and $E^{He}(\rho,T)$
are total energy of He$^{+}$ and He at the relative density and temperature, respectively.

Based on the approximations mentioned above, the calculated EOS was examined
theoretically along the principal Hugoniot, the locus of states that satisfy
the Rankine-Hugoniot (RH) equations, which are derived from conservation of
mass, momentum, and energy across the front of shock waves. The RH equations
describe the locus of states in ($E$, $P$, $V$)-space satisfying the following
relation:
\begin{equation}
\label{equ_hugoniot1}E_{1}-E_{0}=\frac{1}{2}(P_{1}+P_{0})(V-V_{0}),
\end{equation}
\begin{equation}
\label{equ_hugoniot2}P_{1}-P_{0}=\rho_{0}u_{s}u_{p},
\end{equation}
\begin{equation}
\label{equ_hugoniot3}V_{1}=V_{0}(1-u_{p}/u_{s}),
\end{equation}
where $E$, $P$, $V$ present internal energy, pressure, volume, and subscripts
0 and 1 present the initial and shocked state, respectively. In Eqs.
(\ref{equ_hugoniot2}) and (\ref{equ_hugoniot3}), $u_{s}$ is the velocity of
the shock wave and $u_{p}$ corresponds to the mass velocity of the material
behind the shock front. In our present simulations, the initial density for
helium is $\rho_{0}$=0.1233 g/cm$^{3}$ and the liquid specimen is controlled
at a temperature of 4 K, where the internal energy is $E_{0}$=$-$0.02 eV/atom.
The initial pressure $P_{0}$ can be treated approximately as zero compared to
the high pressure of shocked states along the Hugoniot. The Hugoniot points
are obtained as follows: (i) smooth functions are used to fit the internal
energy and pressure in terms of temperature at sampled density; (ii) then,
Hugoniot points are derived from Eq. (\ref{equ_hugoniot1}).

The principal Hugoniot curve for helium is shown in Fig. \ref{fig_hugoniot},
where previous experimental measurements and theoretical predictions are also
plotted for comparison. For pressures below 30 GPa, our results show good
agreements with the data detected by gas gun experiments
\cite{PBX:Nellis:1984}. At higher pressures, the principal Hugoniot curve
shows a maximum compression ratio $\eta_{max}\mathtt{\approx}6$ with the
crossover pressure around 100 GPa, as have been reported by laser driven
experiments \cite{PBX:Eggert:2008}. Soon after that, the use of quartz as a
shock wave standard by Knudson \emph{et al.} has improve the shock data, and
$\eta_{max}$ has been reduced to be 5.1 \cite{PBX:Knudson:2009}. Our corrected
QMD simulation results indicate that, the atomic ionization dominates the
characteristic of the EOS at $P>30$, and the principal Hugoniot shows soften
behavior with the increase of pressure, then reach its maximum ($\eta_{max}%
$=$5.16$) with the corresponding pressure of 106 GPa. Then, stiff behavior has
been found at $P>110$ GPa along the Hugoniot. As has been shown in Fig.
\ref{fig_ion}, with the increase of pressure along the Hugoniot, ionization
energy decreases, and continuous increase of atomic ionization has become
considerable in determining the EOS of fluid helium. Maximum compression ratio
is achieved with the ionization degree of 40\%. The wide-range behavior of our
simulated principal Hugoniot for helium shows excellent agreement with
experimental ones.

\begin{figure}[pt]
\centering
\includegraphics[width=12.0cm]{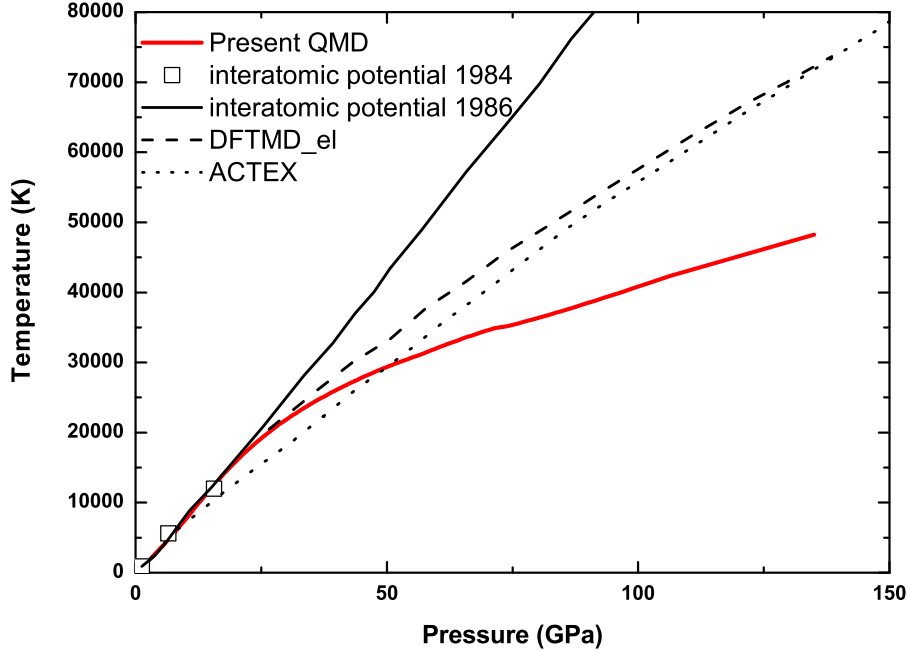}\caption{(Color online) Comparison
between our calculated shock temperature and previous theoretical predictions.
Interatomic potential results are shown as open square \cite{PBX:Nellis:1984}
and solid line \cite{PBX:Ross:1986}; DFT \cite{PBX:Militzer:2006} and ACTEX
\cite{PBX:Ross:2007} results are plotted as dashed and dotted line,
respectively.}%
\label{fig_temp}%
\end{figure}

Concerning other theoretical models, classical MC results show stiff behavior
along the Hugoniot, and $\eta_{max}$ lies around 4 at 1000 GPa
\cite{PBX:Aziz:1995}. Direct DFT simulation results show stiff behavior up to
110 GPa, while, by combining PIMC simulations, $\eta_{max}$ was reported to be
5.3 at 360 GPa \cite{PBX:Militzer:2006}, but the predicted pressure is still
too high compared with laser driven experiments. ACTEX \cite{PBX:Ross:2007}
and SCVH \cite{PBX:Saumon:1995} calculations predict $\eta_{max}$ to be 6 near
100 GPa and 300 GPa respectively, but the results do not agree with those of
Knudson \emph{et al.} \cite{PBX:Knudson:2009}.

Temperature, which is focused as one of the most important parameters in
experiments, is difficult to be measured because of the uncertainty in
determining the optical-intensity loss for ultraviolet part of the spectrum in
adiabatic or isentropic shock compressions, especially for the temperature
exceeding several electron-volt \cite{Chen2010}. QMD simulations provide
powerful tools to predict shock temperature. The calculated Hugoniot
temperature has been shown in Fig. \ref{fig_temp} as a function of pressure
along the Hugoniot. Previous theoretical predictions, such as interatomic
potential \cite{PBX:Nellis:1984,PBX:Ross:1986}, DFT-MD (excited electron)
\cite{PBX:Militzer:2006}, and ACTEX \cite{PBX:Ross:2007} are also provided for
comparison. Disagreements have been found to begin at around 30 GPa with the
relative temperature of 20000 K, which highlights the starting point of
ionization on the Hugoniot, and the predicted temperatures by those models are
higher at a given pressure compared with our calculations.

\section{CONCLUSION}

\label{sec-conclusion}

In summary, We have performed DFT based QMD simulations to study the
thermophysical properties of helium under extreme conditions. The Hugoniot EOS
has been evaluated through QMD calculations and corrected by taking into
account the atomic ionization described by Saha equation, where only first
ionization is considered. The corrected Hugoniot has been proved accord well
with the experimental data in a wide range of shock conditions, which thus
indicates the importance of atomic ionization. Maximum compression ratio of
5.16 reveals at round 106 GPa along the principal Hugoniot, and the softened
characteristic of the Hugoniot has been demonstrated by the contributions from
atomic ionization.

\begin{acknowledgments}
This work was supported by NSFC under Grant No. xxx, by the National
Basic Security Research Program of China, and by the National
High-Tech ICF Committee of China.
\end{acknowledgments}


\begin{thebibliography}{99}                                                                                               %
\bibitem {PBX:Ernstorfer:2009}R. Ernstorfer, M. Harb, C. T. Hebeisen, G.
Sciaini, T. Dartigalongue, and R. J. D. Miller, Science, \textbf{323} 1033 (2009).

\bibitem {PBX:Nellis:2006}W. J. Nellis, Rep. Prog. Phys., \textbf{69} 1479 (2006).

\bibitem {PBX:Hicks:2009}D. G. Hicks, T. R. Boehly, P. M. Celliers, J. H.
Eggert, S. J. Moon, D. D. Meyerhofer,and G. W. Collins, Phys. Rev. B,
\textbf{79} 014112 (2009)

\bibitem {PBX:Philippe:2010}F. Philippe, A. Casner, T. Caillaud, O. Landoas,
M. C. Monteil, S. Liberatore, H. S. Park, P. Amendt, H. Robey, C. Sorce, C. K.
Li, F. Seguin, M. Rosenberg, R. Petrasso, V. Glebov, and C. Stoeckl, Phys.
Rev. Lett., \textbf{104} 035004 (2010).

\bibitem {PBX:Lorenzen:2009}W. Lorenzen, B. Holst, and R. Redmer, Phys. Rev.
Lett., \textbf{102} 115701 (2009).

\bibitem {PBX:Stevenson:1977a}D. J. Stevenson and E. E. Salpeter, Astrophys.
J. Suppl., \textbf{35} 221 (1977).

\bibitem {PBX:Stevenson:1977b}D. J. Stevenson and E. E. Salpeter, Astrophys.
J. Suppl., \textbf{35} 239 (1977).

\bibitem {PBX:Saumon:1995}D. Saumon, G. Chabrier, and H. M. Van Horn,
Astrophys. J. Suppl. Ser., \textbf{99} 713 (1995).

\bibitem {PBX:Ternovoi:2004}V.Ya. Ternovoi \emph{et al.}, JETP Lett.,
\textbf{79} 6 (2004).

\bibitem {PBX:Vorberger:2007}J. Vorberger, I. Tamblyn, B. Militzer, and S. A.
Bonev, Phys. Rev. B., \textbf{75} 024206 (2007).

\bibitem {PBX:Perryman:2000}M. A. C. Perryman, Rep. Prog. Phys., \textbf{63}
1209 (2000).

\bibitem {PBX:Nellis:1984}W. J. Nellis, N. C. Holmes, A. C. Mitchell, R. J.
Trainor, G. K. Governo, M. Ross, and D. A. Young, Phys. Rev. Lett.,
\textbf{53} 1248 (1984).

\bibitem {PBX:Eggert:2008}J. Eggert, S. Brygoo, P. Loubeyre, R. S. McWilliams,
P. M. Celliers, D. G. Hicks, T. R. Boehly, R. Jeanloz, and G.W. Collins, Phys.
Rev. Lett., \textbf{100} 124503 (2008).

\bibitem {PBX:Knudson:2009}M. D. Knudson and M. P. Desjarlais, Phys. Rev.
Lett., \textbf{103} 225501 (2009).

\bibitem {PBX:Ross:1986}M. Ross and D. A. Young, Phys. Lett. A, \textbf{118}
463 (1986).

\bibitem {PBX:Chen:2007}Q. F. Chen, Y. Zhang, L. C. Cai, Y. J. Gu, and F. Q.
Jing, Phys. Plasma., \textbf{14} 012703 (2007).

\bibitem {PBX:Kowalski:2007}P. M. Kowalski, S. Mazevet, D. Saumon, and M.
Challacombe, Phys. Rev. B, \textbf{76} 075112 (2007).

\bibitem {PBX:Militzer:2006}B. Militzer, Phys. Rev. Lett., \textbf{97} 175501 (2006).

\bibitem {PBX:Ross:2007}M. Ross, F. Rogers, N. Winter, and G. Collins, Phys.
Rev. B, \textbf{76} 020502(R) (2007).

\bibitem {PBX:Wang:2010a}C. Wang and P. Zhang, J. Chem. Phys., \textbf{132},
154307 (2010).

\bibitem {PBX:Wang:2010b}C. Wang and P. Zhang, J. Appl. Phys., \textbf{107},
083502 (2010).

\bibitem {PBX:Kresse:1993}G. Kresse and J. Hafner, Phys. Rev. B, \textbf{47},
R558 (1993).

\bibitem {PBX:Kresse:1996}G. Kresse and J. Furthm{\"u}ller, Phys. Rev. B,
\textbf{54}, 11169 (1996).

\bibitem {PBX:Lenosky:2000}T. Lenosky, S. Bickham, J. Kress, and L. Collins,
Phys. Rev. B, \textbf{61}, 1 (2000).

\bibitem {PBX:Perdew:1991}J. P. Perdew, \textit{Electronic Structure of
Solids} (Akademie Verlag, Berlin, 1991).

\bibitem {PBX:Blochl:1994}P. E. Bl{\"o}chl, Phys. Rev. B, \textbf{50}, 17953 (1994).

\bibitem {PBX:Nose:1984}S. Nos\'{e}, J. Chem. Phys., \textbf{81}, 511 (1984).

\bibitem {PBX:timestep}The time steps have been taken as $\vartriangle
t=a/20\sqrt{k_{B}T/m_{He}}$, where $a=(3/4\pi n_{i})^{1/3}$ is the ionic
sphere radius ($n_{i}$ is the ionic number density), $k_{B}T$ presents the
kinetic energy, and $m_{He}$ is the ionic mass.

\bibitem {PBX:Holst:2008}B. Holst,R. Redmer, and M. P. Desjarlais, Phys. Rev.
B, \textbf{77} 184201 (2008).

\bibitem {PBX:Mazevet:2005}S. Mazevet, M. P. Desjarlais, L. A. Collins, J. D.
Kress, and N. H. Magee, Phys. Rev. E, \textbf{71} 016409 (2005).

\bibitem {PBX:Wang:2010c}C. Wang, X. T. He, and P. Zhang, J. Appl. Phys.,
\textbf{108}, 044909 (2010).

\bibitem {PBX:Aziz:1995}R. A. Aziz, A. R. Janzen, and M. R. Moldover, Phys.
Rev. Lett., \textbf{74} 1586 (1995).

\bibitem {Chen2010}Q. F. Chen, private communication.
\end{thebibliography}

\end{document}